\documentclass[f]{ceurart}
\usepackage{listings}
\usepackage{enumitem}
\usepackage{xcolor}

\begin{document}
\copyrightyear{2024}
\copyrightclause{Copyright for this paper by its authors.
  Use permitted under Creative Commons License Attribution 4.0
  International (CC BY 4.0).}
\conference{Under review for CLAIRvoyant - ConventicLE on Artificial Intelligence Regulation - 2024 @ JURIX 2024}
\title{Towards An Automated AI Act FRIA Tool That Can Reuse GDPR's DPIA}
\author{Harshvardhan J. Pandit}
\cormark[1]
\author{Tytti Rintamäki}
\cormark[1]
\address{ADAPT Centre, School of Computing, Dublin City University, Dublin, Ireland}
\cortext[1]{Corresponding author: Tytti Rintamäki tytti.rintamaki@adaptcentre.ie and Harshvardhan J. Pandit me@harshp.com}

\begin{abstract}
The AI Act introduces the obligation to conduct a Fundamental Rights Impact Assessment (FRIA), with the possibility to reuse a Data Protection Impact Assessment (DPIA), and requires the EU Commission to create of an automated tool to support the FRIA process. In this article, we provide our novel exploration of the DPIA and FRIA as information processes to enable the creation of automated tools. We first investigate the information involved in DPIA and FRIA, and then use this to align the two to state where a DPIA can be reused in a FRIA. We then present the FRIA as a 5-step process and discuss the role of an automated tool for each step. Our work provides the necessary foundation for creating and managing information for FRIA and supporting it through an automated tool as required by the AI Act.
\end{abstract}
\textcolor{red}{Presented at CLAIRvoyant (ConventicLE on Artificial Intelligence Regulation) Workshop 2024}

\maketitle

\section{Introduction}
The rapid proliferation of artificial intelligence (AI) technologies presents novel challenges for creating new regulations to guide responsible progress and prevent harmful impacts. The EU's recently published AI Act \cite{EU_AIAct}, which is the world's first legal framework to regulate AI,  tackles this challenge through a risk based approach where AI systems are classified as `\textit{high-risk}' based on their potential impact to fundamental rights. To guide organisations and authorities in detecting where such impacts might arise, the AI Act's Article 27 lays down the obligation for high-risk AI systems to conduct a `Fundamental Rights Impact Assessment' (FRIA) that identifies ``risks of harm likely to have an impact'' on people or groups of people affected by the use of the AI system. 

This framework of using impact assessments is based on and similar to the EU's General Data Protection Regulation (GDPR) \cite{EU_GDPR}, a globally impactful regulation that regulates the processing of personal data and includes an impact assessment obligation called the `Data Protection Impact Assessment' (DPIA). Under Article 35 of the GDPR, organisations are required to conduct a DPIA when their data processing activities are likely to result in a high risk to the rights and freedoms of individuals. This requirement to conduct a DPIA can be based on the list of use-cases defined in GDPR, or through lists published by Data Protection Authorities (DPAs) responsible for enforcing the GDPR. 

In our prior work \cite{rintamakiHighRiskCategorisationsGDPR2024}, we found there are 94 distinct conditions for which a DPIA can be required based on the GDPR or the country involved, and that of the 25 high-risk conditions in AI Act's Annex III, GDPR applies and a DPIA is necessary in 22 clauses, and could conditionally be required in the remaining 3. Thus there is an inherent synergy between the GDPR and the AI Act. While there will be use-cases where AI systems do not involve personal data, there will also be a large number of AI systems using personal data - either as inputs when developing the underlying AI models and components or as inputs to the system and by producing outputs that can be considered as personal data. This necessitates a combined implementation of GDPR and the AI Act.

These findings reveal the wide overlap between GDPR and AI Act and establish the importance of exploring the synergy between their respective impact assessments, namely the DPIA in GDPR and the FRIA in AI Act as both require the assessment of risks to rights and freedoms and utilising the outputs to avoid harms to people. The drafting of the AI Act indeed acknowledges this by stating in Art.27-5 that ``If any of the obligations ... is already met through the DPIA conducted pursuant to Article 35 of GDPR, the FRIA ... shall complement that DPIA''. However, the AI Act does not expand upon this statement in any other article or recital, thereby raising the question of how and where can a DPIA be used with a FRIA.

In addition to this, the AI Act's Article 27-5 states ``The AI Office shall develop a template for a questionnaire, including through an automated tool, to facilitate deployers in complying with their obligations under this Article in a simplified manner''. This means the enforcement of the AI Act requires creating an information system that will assist with the obligations associated with FRIA, and that such an information system will use automation to simplify and support stakeholders in their compliance related activities. In the past, similar approaches have existed for conducting DPIA under the GDPR - though in the form of guidance from DPAs. For example, CNIL, the French DPA, has a tool that supports DPIA processes by utilising a risk assessment approach to collecting and documenting information \cite{OpenSourcePIAa}.

While this is a progressive and positive step in terms of regulations acknowledging and requiring the use of technologies to aid in simplifying and enforcing compliance, the implications of the AI Act in this regard are not completely clear. For example, there are several possibilities for what the ``automated tool'' might refer to - it could be a simple checklist that ascertains where and when a FRIA is needed, or it could be a tool to conduct the FRIA itself, or it could be a tool to support the notification of the FRIA to stakeholders (including authorities). There are also necessary questions that must be raised as to how such a tool can be made to support the variety and diversity of use-cases that exist, and how its information can be made to be interoperable so that other tooling can be used to address tasks not covered by the tool. And how such a tool, if it is intended to, facilitate the reuse of a DPIA for FRIA.

Using this as the motivation, we outline the following research questions:
\setlist{nolistsep}
\begin{enumerate}[label=\textbf{RQ\arabic*},noitemsep]
    \item How to interpret DPIA obligations in a manner that facilitates its reuse for a FRIA (Section 3)
    \item Where can automation assist with FRIA obligations? (Section 4)
\end{enumerate}

By exploring these questions, we provide the first discussion on the implementation of Article 27 of the AI Act regarding how to approach the FRIA from an information systems and knowledge engineering perspective, the intersection and reuse of GDPR's DPIA with the FRIA, and the implications of these on the creation and use of automated tools to support both DPIA and FRIA processes. Our work has important implications for all stakeholders - for AI authorities and organisations using AI it provides a framework for how to approach the implementation of FRIA, how to interpret the FRIA information in a manner that aids in the creation of automated tooling for different use-cases, and how the DPIA can be integrated in various ways to support the FRIA. 

For the GDPR authorities and organisations processing personal data, our work provides a way to adapt existing DPIA processes and tooling to address the additional requirements of a FRIA. Our work also has implications on the combined enforcement of GDPR and AI Act and the co-operation of their respective authorities as it outlines what informational overlaps exist in the obligations for DPIA and FRIA, and how these can be combined and used to support common obligations.

\section{State of the Art}

\subsection{AI Act and FRIA}
The AI Act and the FRIA are both recent developments. Therefore the state of the art on this topic is minimal, though it is expected to be a growing discussion as the AI Act starts its enforcement activities. Mantelero \cite{manteleroFundamentalRightsImpact2024} discusses the AI Act and FRIA as a legal processes and compares it to the approaches taken by GDPR's DPAs regarding the use of AI. Calvi and Kotzinos \cite{calviEnhancingAIFairness2023} discuss the FRIA as part of a wider analysis of impact assessments across GDPR, AI Act, and Digital Services Act (DSA). Similarly, Malgieri and Santos \cite{malgieriAssessingSeverityImpacts2024} also analyse impact assessments across these regulations and proposal a rights-based approach to assess impacts through a multi-metric representation of severity. These approaches provide clarity on the legal processes involved in FRIA and how these interact with the GDPR.

The most prominent early work regarding FRIA was the ALIGNER Fundamental Rights Impact Assessment (AFRIA) \cite{FundamentalRightsImpact}, produced as part of the ALIGNER H2020 project to support responsible use of AI in law enforcement activities. The AFRIA is a structured template for a FRIA based on evaluating the occurrence and severity of specific issues, based on which a risk level is identified. Golpayegani et. al. \cite{golpayeganiAICardsApplied2024} propose the creation of `AI Cards' as a visual and machine-readable framework for representing information based on the AI Act, and which can be used to support the FRIA process. They also propose the interpretation of changes to intended purpose based on the combination of specific granular concepts. 

Janssen et. al. \cite{janssenPracticalFundamentalRights2022} describe FRIA as a four-step process comprised of (i) defining the system's purposes and responsibilities of entities involved (ii) assessing risks for system development; (iii) justifying why risks on rights are proportionate; and (iv) identifying organisational and technical measures to mitigate risks. Inverardi et. al. \cite{inverardiFundamentalRightsAI2024} propose calculating a FRIA score that measures the potential impact on fundamental rights based on an open-ended survey for collecting information and using a quantitative matrix to measure the impact on each fundamental right.

We use these approaches to inform our exercise of determining information involved in a FRIA process in Section 3, and to depict the FRIA as a 5-step process that encompasses the broader obligations regarding determining FRIA necessity and notifications in Section 4.

\subsection{GDPR and DPIA}
The GDPR and DPIA are by comparison a well established process with several methodologies and technological tools being developed and used to assist with the relevant legal and organisational processes \cite{georgiadisPrivacyImpactAssessment2022}. We used these approaches to inform our understanding of the information involved in a DPIA process, in particular for how to align its rights impacts with a FRIA in Section 3 and its reuse in Section 4. 

The Danish Institute for Human Rights has released a Human Rights Compliance Assessment tool \cite{thedanishinstituteforhumanrightsHUMANRIGHTSIMPACT2020} which builds upon the DPIA process with a focus on human rights. 
The Fundamental Rights and Algorithms Impact Assessment (FRAIA) approach created by the Dutch government \cite{gerardsFundamentalRightsAlgorithms2022} is based on adapting the DPIA process for a broader assessment of rights impacts, though it was finalised before the AI Act. Similarly, Janssen proposed a FRIA framework approach for the private sector based on DPIAs \cite{janssenApproachFundamentalRights2020a}. Cobbe et. al. \cite{cobbeArtificialIntelligenceService2021} explore the implications of GDPR's roles with those defined in the AI Act.
Pandit \cite{panditSemanticSpecificationData2022} expressed the DPIA as an information process that has multiple stages and where the information involved can be expressed in machine-readable form to create automated tools. 

\section{How to use a DPIA for a FRIA?}
To answer the question of how to interpret DPIA obligations in a manner that facilitates its reuse for a FRIA (RQ1), with the eventual goal of creating an automated tool (RQ2 and RQ3), we first establish in Section 3.1 what information is involved in a DPIA. We perform a similar exercise in Section 3.2 regarding the FRIA process. In Section 3.3 we explore the different roles a DPIA can take within the FRIA process - namely as an ex-ante input where an existing DPIA supports a new FRIA, or as an ex-post activity where both the DPIA and the FRIA are conducted simultaneously. We then use these to align the DPIA process with the FRIA process in Section 3.4 based on the information involved. 

\subsection{Information Requirements for DPIA}
A DPIA under GDPR is an ex-ante activity that is carried out before the processing of personal data to ascertain whether the processing is ``likely to result in a high risk to the rights and freedoms of natural persons'' (GDPR Art.35). The first step in the DPIA process is therefore ascertaining whether a DPIA is required. The GDPR states that assessing whether a DPIA is required should be done ``in particular (when) using new technologies'' - which Artificial Intelligence (AI) technologies are. The GDPR also requires a DPIA where automated processing produces `legal or similarly significant effects' (Art.35-3a), where large scale special categories of personal data are involved (Art.35-3b), and where there is a systematic large scale monitoring of a public area (Art.35-3c). 

In addition to these, GDPR allows enforcing DPAs to create lists of processing activities for which a DPIA is required (Art.35-4) - thereby enabling each of the 30 EU/EEA DPAs to add additional criteria for where a DPIA is needed. In prior work \cite{rintamakiHighRiskCategorisationsGDPR2024}, we analysed these 30 lists along with the GDPR and the EDPB guidelines and found 94 unique conditions for when a DPIA is needed based on the jurisdiction involved. This establishes that the existence of a DPIA is dependent on the jurisdiction involved and the activity being performed, and that the same activities may or may not have a DPIA based on the jurisdictions - unless a DPIA is required under GDPR or through EDPB guidelines which are binding on all 30 EU/EEA jurisdictions. GDPR also allows DPAs to create lists of processing activities which don't require a DPIA (Art.35-5) which represent exceptions for those specific jurisdictions. For the purposes of this work, such exceptions are out of scope at this point.

Once conducting a DPIA is identified as being necessary, the second step in the DPIA process is to identify and collect the relevant information needed to identify the ``risks to the rights and freedoms of data subjects'' (Art.35-7c) and ``measures envisaged to address the risks'' (Art.35-7d). The `rights and freedoms' mentioned in GDPR are all-encompassing i.e. they refer to the fundamental rights and freedoms established in the EU as well as specific rights such as those defined in the GDPR. The `measures' identified in response to risks are also required to take in to account these rights i.e. the measures should not merely satisfy a technical vulnerability or threat, but should also be ensured to be compatible or in balance with the very rights and freedoms they are aimed to protect.

The relevant information required to conduct a DPIA involves ``a systematic description of processing operations and the purposes of processing'' (Art.35-7a), the legal basis used to justify it - in particularly when it is legitimate interests (Art.35-7a), and an ``assessment of the necessity and proportionality of the processing operations in relation to the purposes'' (Art.35-7b). While the GDPR does not explain what `systematic description' means, we can interpret this information from the other articles and their respective obligations, in particular those associated with rights established in Articles 12 to 23, the technical and organisational measures or safeguards mentioned in Article 24, 25, 32 and 33, and the information required to be documented in Articles 30 and 36. Based on this, we establish the following information requirements as inputs to the DPIA process:
\setlist{nolistsep}
\begin{enumerate}
    \item Purposes of processing;
    \item Processing operations;
    \item Personal data involved - and whether it belongs to a special category;
    \item Data subjects - and whether they belong to a vulnerable category;
    \item Entities and their roles e.g. controllers and recipients;
    \item Cross-border transfer of personal data;
    \item Scale of data, processing operations, and data subjects;
    \item Duration of processing operations - including data storage and deletion periods;
    \item Legal bases used to justify the processing - in particular consent and legitimate interests;
    \item Necessity of processing and data e.g. as a `statutory or contractual requirement' (Art.13-3e), which is defined by the EDPS as ``a combined, fact-based assessment of the effectiveness of the measure for the objective pursued and of whether it is less intrusive compared to other options for achieving the same goal'' \cite{NecessityToolkitEuropean2024};
    \item Proportionality of processing and data in relation to the purpose, which is defined by the EDPS as ``both the necessity and the appropriateness, ... that is, the extent to which there is a logical link between the measure and the (legitimate) objective pursued'' \cite{EDPSGuidelinesAssessing2024};
    \item Profiling, Scoring, and Decision making in processing;
    \item Automation in processing;
    \item Inferences or Predictions e.g. ``performance at work, economic situation, health, personal preferences or interests, reliability or behaviour, location or movements'' (Recital 75);
    \item Technical measures used to safeguard the data e.g. pseudonymisation;
    \item Organisational measures (including legal measures) e.g. policies;
    \item Quality of data - in particular its accuracy and completeness; and
    \item Locality of processing e.g. public areas, personal or work places.
\end{enumerate}

After identifying this information, the next step in the DPIA processing is to perform a risk assessment to identify the risks to rights and freedoms (Art.35-7c). The guidelines on DPIA authored by the Article 29 Working Party and endorsed by the EDPB state the following should be identified for each risk:
\setlist{nolistsep}
\begin{enumerate}
    \item Risks sources (Recital 90);
    \item Threats that could lead to the risk;
    \item Likelihood and Severity of the risk (recital 90);
    \item Impacts to the rights and freedoms of data subjects;
\end{enumerate}

This process can be described as identifying a particular \textit{operational risk}, such as ``illegitimate access, undesired modification, and disappearance of data'', for which the risk sources and threats are identified as causing the risk, and the potential impacts are identified as the consequences of the risk. Based on the severity of the impacts and the likelihood of risk events, a risk level is calculated that determines whether the activity is `\textit{high risk}'. 

Following this, the DPIA process requires identifying appropriate measures to address the identified risks (Art.35-7d). Though the GDPR does not dictate what such measures are, the existing literature on risk assessments such as the ISO 31000 series have established different `risk treatment strategies' to manage risks e.g. identification, elimination, substitution, and reduction of likelihood or severity associated with events. These enable actions such as identification of potential risk sources and threats and removing them or reducing the likelihood of vulnerabilities that could give rise to risks, as well as identifying the consequences of risks and eliminating them or reducing the likelihood or severity of their impact. In addition to these, GDPR Art.32 also describes specific measures such as pseudonymisation and encryption, ensuring ongoing confidentiality, integrity, availability, and resilience of processing systems and services, ability to restore the availability and access to personal data, and regularly testing, assessing and evaluating the effectiveness of technical and organisational measures.

The final step in a DPIA process is to assess the residual risk and determine whether the processing can be (safely) carried out. For processing that is found to be high-risk without any mitigation measures, the GDPR requires consulting a DPA with a completed DPIA (Art.36-3), and based on which the DPA may require ceasing the processing operations (Art.58-2f). The DPIA thus represents an approval process through which organisations identify (high-)risks to rights and freedoms, communicate these to the DPA along with measures taken to safeguard against impacts, and obtain authorisation to continue the proposed processing operations.

\subsection{Information Requirements for FRIA}
A FRIA under the AI Act is also an ex-ante activity that is carried out before deploying a high-risk AI system as determined under the conditions outlined in Art.6-2, and is intended to be ``an assessment of the impact on fundamental rights that the use of such system may produce'' (Art.27-1). Similar to the DPIA process, the first step in the FRIA process is to assess whether a FRIA is required based on the classification of an AI system as high-risk, and the obligations and exceptions established in Articles 6 and 27. For the scope of this work, we only focus on the cases where a FRIA is necessary and do not discuss the exceptions for entities and use-cases that do not require a FRIA. 

Once the obligation to conduct a FRIA has been established, the second step in the FRIA process is to identify and collect the relevant information to carry out the impact assessment on fundamental rights - which the AI Act states as being ``health, safety, fundamental rights as enshrined in the Charter of Fundamental Rights of the European Union'' (Recital 1). And where the `risks' as per Art.27-1d are specific to the ``risks of harm likely to have an impact on the categories of natural persons or groups of persons``. Thus the broader interpretation of the FRIA is to assess the risk of harmful impacts to fundamental rights as enshrined within the Charter.

The relevant information required to conduct a FRIA involves ``a description of the deployer's processes in which the AI system will be used in line with its intended purpose'' (Art.27-1a), the duration and frequency of use (Art.27-1b), and the categories of natural persons and groups likely to be affected by use (Art.27-1c). While the AI Act does define `intended purpose' (Art.3-12) as ``the use for which an AI system is intended by the provider, including the specific context and conditions of use, as specified in the information supplied by the provider in the instructions for use, promotional or sales materials and statements, as well as in the technical documentation'', it is open to interpretation what information this precisely includes. Similarly, the AI Act does not clarify what `description of the deployer's process' and `affected by use' mean. We therefore interpret this information from the other articles and their respective obligations - in particular those regarding data and data governance (Art.10), technical documentation (Art.11 and Annex IV), record keeping (Art.12), instructions of use (Art.13), and transparency (Art.50). Based on this, we establish the following information requirements as inputs to the FRIA process by reusing our developed understanding of how to describe the use of personal processing for a DPIA in Section 3.1:
\setlist{nolistsep}
\begin{enumerate}
    \item Intended Purpose(s) as defined in instructions of use, technical documentation, and promotional and sales materials, and interpreted as the combination of purpose(s), AI capabilities, and domain(s) as per \cite{golpayeganiAICardsApplied2024}, where purposes are defined as:
    \begin{enumerate}
        \item Purpose(s) for which an AI system was developed (Art.3-3);
        \item Purpose(s) for which the AI system was put on the market (Art.3-9);
        \item Purpose(s) of data collection for AI systems developed and operated with personal data (Art.10-2b);
    \end{enumerate}
    \item Involved Entities consisting of:
    \begin{enumerate}
        \item Entities in the role of deployers and providers, and their relation and control regarding the AI system e.g. capacity to update/change (Art.13-3c);
        \item Entities acting as `users' of the AI system e.g. an immigration officer overseeing an automated passport control service;
        \item Entities subjected to the use of AI system e.g. a passenger passing through the automated passport control service (hereafter referred to as `AI subject' based on the similar `data subject' under GDPR);
        \item along with the Context of AI subjects' interactions with the AI system - whether they are active/passive, intended/unintended, and informed/uninformed \cite{golpayeganiAICardsApplied2024};
        \item the capacity of users and AI subjects to control the AI system e.g. verify or change the outputs \cite{golpayeganiAICardsApplied2024};
    \end{enumerate}
    \item Involved Data consisting of:
    \begin{enumerate}
        \item Input data, including personal data (Art.10-2b and Art.13-3b-vi)
        \item Processing of data (Art.10-2c) where `processing' includes operations defined in GDPR Art.4-2;
        \item Output data, including personal data, and whether it can be classified as profiling and decisions;
        \item Data quality (Art.10-2e)
        \item Whether personal data involved is of Special Category under GDPR Article 9 (AI Act Art.10-5);
    \end{enumerate}
    \item Deployment information consisting of:
    \begin{enumerate}
        \item Interactions of the AI system with other systems and components (Anx.IV-1b);
        \item Modalities or forms under which the AI system is to be provided and used (Anx.IV-1d);
        \item Hardware and software requirements (Art.13e and Anx.IV-1e);
        \item User-interface used to operate and manage the AI system (Anx.IV-1g);
        \item Duration and Frequency for use of AI system;
    \end{enumerate}
    \item Provenance of AI system consisting of:
    \begin{enumerate}
        \item How the AI system was developed, including the resources used to develop, train, test and validate the AI system and their provenance (Anx.IV-2a and Anx.IV-2c);
        \item `Datasheets' describing the training methodologies and techniques and the training data sets used, their provenance, scope and main characteristics; data collection and labelling/cleaning procedures (Anx.IV-2d);
        \item Relevant changes made by the provider to the system through its lifecycle (Anx.IV-6)
    \end{enumerate}
    \item Operational information consisting of:
    \begin{enumerate}
        \item Expected outputs (Anx.IV-2b);
        \item Logic of the AI system and algorithms, including rationale and assumptions, optimisations, trade-offs and `output quality' (Anx.IV-2b);
        \item Pre-determined changes to the AI system and its performance (Art.13-3c and Anx.IV-2f)
        \item Appropriateness of the performance metrics (Anx.IV-4);
    \end{enumerate}
\end{enumerate}

Based on this information, the next step in the FRIA process is to identify the risks of harm to ``natural persons and groups likely to be affected by its use in the specific context'' (Art.27-1c). Here, `affected' can be interpreted to be limited to negative effects in the context of the FRIA's goals, and taken to refer to events such as various kinds of denial, detriment, damage, harm, discrimination, fraud, and loss which can be in the context of the AI system or service (e.g. denial of service), a physical activity (e.g. physical harm), or psychological or social implication (e.g. discrimination at the workplace). The goal of the FRIA process is thus to identify whether such `effects' can be considered as `risks of harm' and as `impacts' to the fundamental rights and freedoms.

The AI Act lays down a detailed risk management procedure in Art.9 which first requires identification and analysis of risks that are known based on the use of the AI system as per the intended purpose, or under `reasonable foreseeable misuse' (Art.9-2b), and through analysis of post-marketing monitoring system (Art.9-2c). The AI Act itself refers to several risks, risk sources, and threats, such as: risks of harm based on information given by provider (Art.27-1d), biases (Art.10-2e), issues with metrics such as accuracy (Art.13-3b), robustness (Art.13-3b), cybersecurity (Art.13-3b), and `reasonably foreseeable misuse' which is ``the use of an AI system in a way that is not in accordance with its intended purpose, but which may result from reasonably foreseeable human behaviour or interaction with other systems, including other AI systems'' (Art.3-13). The AI Act also refers to operational risks such as ``capabilities and limitations in performance such as accuracy for specific people and applications and the overall expected level of accuracy in relation to its intended purpose'' (Art.13-3b-ii, Art.13-3b-v,  and Anx.IV-3). For each risk and impact identified, the AI Act requires assessing its probability (likelihood) and severity (Art.3-2) which are used to determine ``foreseeable unintended outcomes and sources of risks to health and safety, fundamental rights and discrimination'' (Art.9-2a, Art.13-3b-iii, Anx.IV-3).

Identified risks are to be treated with specific and relevant measures, which can be elimination or reduction (Art.9-5a), or mitigation (Art.9-5b), and can consist of technical measures such as controls in/over the AI system or organisational measures such as awareness and training. The AI Act mentions specific measures such as bias mitigation (Art.10-2f), validation and testing data (Anx.IV-2g), assessing metrics such as accuracy, robustness, and cybersecurity (Art.13-3b and Anx.IV-2g), human oversight (Art.27-1e), and maintenance procedures (Art.13-3e). The AI Act also mentions maintaining and analysing logs to support the risk management process (Art.12 and Art.13-3f), using technical measures to facilitate the interpretation of outputs (Art.13-3b and Anx.IV-3d). For FRIA, such measures also include governance procedures and complaint mechanisms (Art.27-1f), 

The final step in the FRIA process is to assess the residual risk and determine whether the AI system can be (safely) deployed, and based on the applicable obligations - either perform a self-assessment (Art.6-3) or to notify the Market Surveillance Authority (Art.27-3). The FRIA thus represents a process through which deployers can assess the safety in relation to (high-)risks to fundamental rights for their AI system before and after putting it on the market, and where applicable, obtain authorisation to proceed with the deployment.

\subsection{DPIA as an Ex-Ante and Ex-Post input to FRIA}
In the previous two sections, we explored the DPIA and FRIA processes in terms of the information required to conduct them as well as the outputs produced and their use with in the respective scopes of GDPR and AI Act. In this section, we discuss where a DPIA may be used with FRIA based on the AI Act's statements regarding the existence of prior impact assessments in Art.27-2 and the functioning of the AI value chain in Art.25. 

The phrasing of AI Act's Art.27-4 regarding FRIA as ``If any of the obligations laid down in this Article is already met through the DPIA ... the FRIA shall complement that DPIA'' implies the use-case where a completed DPIA can be used in a manner that satisfies some of the obligations of a FRIA and therefore reduce the requirements for conducting a FRIA. We refer to this use-case as DPIA being an `\textit{ex-ante}' input in order to start a FRIA. In order to enable such reuse of a DPIA, we need to first understand the respective obligations of DPIA and FRIA, where they overlap, and which obligations of a FRIA do not need to be repeated if they have already been achieved through the DPIA.
From an information systems perspective, with the intended goal of building an automated tool, this can be formulated as the following research question: ``\textit{To what extent can information and outputs from a DPIA process be reused as inputs in a FRIA process? Based on this, what additional information is required to complete the FRIA process when reusing an existing DPIA?}'' 

The second use-case interprets the phrasing in Art.27-4 regarding `complimenting a FRIA with a DPIA' as referring to the two activities occurring simultaneously in a complimentary manner. This is a novel but necessary interpretation as even though the AI Act does not dictate how its obligations function with GDPR, our prior work has shown that most of the high-risk cases defined in AI Act's Annex III require a DPIA \cite{rintamakiHighRiskCategorisationsGDPR2024}. Thus deployers, where they qualify as controllers under the GDPR, have to conduct both a DPIA and a FRIA for an AI system in order to satisfy both GDPR and AI Act obligations. Rather than first conducting a DPIA and then using it to conduct a FRIA, both DPIA and FRIA processes can be carried out simultaneously based on the similarities in their scopes and objectives. We refer to this use-case as a DPIA being an `\textit{ex-post}' output after starting a FRIA. From an information systems perspective, with the intended goal of building an automated tool, we formulate this use-case as the following research question: ``\textit{To what extent can a DPIA and a FRIA be conducted simultaneously based on overlaps in the information required? Based on this, what additional information is required to complete the DPIA and FRIA processes?}''.

\subsection{Reuse of DPIA Information in FRIA}
For both cases whether the DPIA is an ex-ante input to the FRIA or an ex-post output, as outlined in the previous section, we need to identify the commonality in information requirements for both so as to build technical tools that can support the completion of FRIA obligations based on information present in/through a DPIA. To do this, we take the information concepts identified in Section 3.1 regarding DPIA and Section 3.2 regarding FRIA and create a mapping between them to identify similarities and overlaps. To start with, we first align the clauses in GDPR Art.35 and AI Act Art.27 to understand the commonality in legal obligations, and then explore the alignment in information required to meet each obligation. A detailed exploration of such an alignment between the obligations and information across both regulations requires further work and is beyond the scope of this current article (we explicitly mention this later as future work).

\noindent\textbf{Systematic Description:} GDPR Art.35-7a and AI Act. 27-1a and Art.27-1b both require a systematic description of the respective systems as inputs to the impact assessment process. This includes purposes, technical operations, involved data, human (data or AI) subjects, and context such as scale of data and operations, cross-border implications, controls for entities (e.g. consent, oversight), whether the activities involve profiling, scoring, or decision making, and the environment (e.g. public area).

By comparing the systematic descriptions required for a DPIA (Section 3.1) with that for a FRIA (Section 3.2), we find that both require identifying common elements such as the purposes, data - and whether it is personal and special category under GDPR, human subjects (whether as data subjects or AI subjects), technical descriptions of activities such as the processing operations over data, scale and scope of data, processing, and subjects, location of operations, contextual considerations such as the involvement of profiling and decision making, the specific technical and organisational measures used, data quality and governance measures implemented, and the `necessity and proportionality' of the operations against the (intended) purposes.

We also find that there are some items which do not share an overlap - such as the interpretation of legal bases in the DPIA process under GDPR, or the involvement of non-personal data under the AI Act. Further, the AI Act has more information requirements which do not correspond to requirements for the DPIA - such as the specific use of data in training and validation, user interfaces to manage the AI systems and explain its outputs, provenance of the AI system and its development, and planned changes to the AI system. 

However, where these concepts involve personal data, they become relevant for a DPIA as part of describing the processing operations (e.g. personal data used to train AI systems) or as part of specific measures (e.g. explanation of AI outputs can be useful as a measure for automated processing). Thus, systematic description of an AI system as required in a FRIA can also aid and compliment the DPIA process. This means both ex-ante and ex-post uses of DPIA as described in the previous section are feasible and can be facilitated through the same automated tool being built to support the FRIA process. Conversely, this also means existing DPIA tools can be adapted to support the FRIA process.

\noindent\textbf{Proportionality and Necessity:} Though the AI Act does not directly use these terms, these concepts are well established in EU law \cite{EDPSGuidelinesAssessing2024} and are a necessary tool in impacts assessments. For GDPR, proportionality and necessity (GDPR Art.35-7b) refer to the processing of personal data for specific purposes. For AI Act, proportionality and necessity can be interpreted to refer to the use of AI systems to achieve the intended purpose (AI Act Art.27-1a) - a crucial concept which is the basis for interpretation of other important concepts such as reasonably foreseeable misuse (Art.3-13), instructions for use (Art.3-14), and performance of an AI system (Art.3-18). Where AI systems involve the processing of personal data, there is therefore an overlap in the assessment of proportionality and necessity assessments as both GDPR and the AI Act assess this based on their respectively defined purposes.

\noindent\textbf{Risks to Rights and Freedoms:} GDPR's DPIA refers to ``risks to the rights and freedoms of data subjects'' (Art.35-7c), while the AI Act's FRIA refers to ``assessment of the impact on fundamental rights'' (Art.27-1d) and ``significant risk of harm to the health, safety or fundamental rights of natural persons'' (Art.6-3). The DPIA thus has a broader scope regarding impacts to rights as it includes both fundamental and other rights, whereas the FRIA specifically focuses on fundamental rights. Both DPIA and FRIA require an assessment of likelihood and severity of risks and use this information to identify a risk level to decide on further obligations. Further, both require an assessment of whether risks are likely to impact rights (in the case of AI Act - fundamental rights). 

While GDPR does not explicitly describe how to assess `impacts', it defines `material or non-material damage' (Art.82) as a criteria for compensation. The AI Act uses the phrasing `risk of harms' to narrow the categories of impacts to events that can be considered as `harms' under EU law, and also provides explicit examples including damage and disruption (Recital 155). Thus, while the initial process of risk assessment is similar in both DPIA and FRIA, the assessment and expression of impacts have differing legal terminology and criteria that depend on their respective regulations.

\noindent\textbf{Risk Mitigation Measures:} Both DPIA (GDPR Art.35-7d) and FRIA (AI Act Art.27-1f) require identifying specific measures in response to risk through risk assessment and management processes. While the GDPR includes specific technical and organisational measures (Art.32), these are focused on confidentiality, integrity, availability regarding the processing of personal data regarding (Art.35-7d). In contrast, the AI Act provides a detailed framework consisting of risk management (Art.9), data governance practices (Art.10), technical documentation (Art.11, Anx.IV), human oversight (Art.14), ensuring accuracy, robustness, and cybersecurity (Art.15), and establishing a quality management system (Art.17).

From this, we can see that the DPIA's notion of risk mitigation measures is limited as compared to the AI Act, though both have the same ultimate goal of using these measures to mitigate risks regarding impacts on rights. Based on where and how an AI system involves and processes data, and the interpretation of where the scope of GDPR starts in this process, the measures described in AI act also have a bearing on the implementation of the GDPR as they will be considered as technical and organisational measures that safeguard the processing of personal data and address risks to rights. Similarly, the specific technical and organisational measures intended to safeguard the processing of personal data, and used within a DPIA, would also be relevant to support the AI Act's obligations regarding risk management based on the involvement of personal data within the AI system and its lifecycle.

\section{How to use an `Automated Tool' in FRIA?}

\subsection{FRIA as a 5-step process}
The AI Act Art.27-5 requires the AI Office to develop a questionnaire and an automated tool to ``to facilitate deployers in complying with their obligations ... in a simplified manner''. Recital 96 further clarifies that this requires the AI Office to ``develop a template for a questionnaire in order to facilitate compliance and reduce the administrative burden for deployers''. From this, it is clear that the questionnaire is to be filled in by the deployer (interpreting the term `template'), and that the outputs of the questionnaire are to support obligations outlined in the AI Act, with the automated tool reducing the efforts involved in some form. 
Further ahead in this section, we outline our interpretation of stages in the FRIA process along with an overview for what information is involved and how the questionnaire and tool could be used for that stage.

However, in order to discuss tools from an information system and software development perspective, the information provided in the AI Act is vague and not sufficient for creating a tool. For example, what does ``facilitate with obligations in simplified manner'' (Art.27-5) mean? Based on the prior establish reuse of privacy engineering techniques to address DPIA and other GDPR compliance processes where tools not only support in meeting legal obligations but also aid organisations in integrating the processes internally and managing the associated information \cite{novelliAutomatingBusinessProcess2023}, and an interpretation of the DPIA as a multi-stage process \cite{panditSemanticSpecificationData2022}, we interpret the FRIA as a 5 stage process where the questionnaire and tool (hereafter we refer to both as just the `tool') can address a specific stage or multiple successive stages.

In Section 3.2 we identified the information required in the FRIA process, and in Section 3.4 we organised them in 4 broad categories as: systematic description, proportionality and necessity assessment, risks to rights, and risk mitigation. These 4 categories, in the specified order, provide a structured approach to collecting and using the information within a FRIA process. To complete the obligations in Art.27-3 regarding deployers notifying market surveillance authorities the FRIA outputs through the filled-out template, we add a fifth category of information corresponding to the last stage in Section 4.1 regarding notifying authorities. The FRIA questionnaire and automated tool can thus work in the following manner for each stage.

\subsection{Stage 1: Determining FRIA necessity}
In the first stage, the tool helps decide whether a FRIA will be needed. The obligation a tool can help with here is Art.27-1 and Art.6-2 in which an entity determines whether it is required to conduct a FRIA based on its role under the AI Act (deployer or provider), the categorisation of the AI system as high-risk (Art.6, Anx.I, and Anx.III), and whether there is a valid exception or exemption for its use-case (Art.6 and Art.27). The tool can vary from being a completely manual process that only aids in identifying the requirements to assess whether a DPIA is required (e.g. a checklist) to being a completely automated process where a decision for whether a FRIA is needed is provided as an output.

In this stage, The tool takes as input information regarding the AI system being assessed, such as whether it is being developed and put on the market and/or it is being deployed, whether it satisfies any high-risk criteria established in Art.6 or Art.27, and whether it is subject to any exception or exemption regarding the classification of high-risk and the requirement for a FRIA. If the tool is intended to support such classifications through automation, it can utilise existing approaches which take specific information categories and determine the risk level based on logical reasoning \cite{golpayeganiAICardsApplied2024}. The final output of the tool is a binary result that indicates whether a FRIA is necessary or not.

We do not provide a detailed description of the information required for the tool in this first stage associated with deciding whether a FRIA is needed as this requires a lengthy analysis of the AI Act as a whole to define the role of entities and the classification of high-risk in Annex I, but we assume it is a pre-requisite step that must be completed before any of the other stages are initiated.

\subsection{Stage 2: Reusing DPIAs}
The second stage is an optional process where the tool helps reuse an existing DPIA as outlined in Section 3.3. The inputs for this process will be a completed DPIA, and the tool will help with obligations in Art.27-2 and Art.27-4. Its outputs consist of producing the information in Section 3.2, and integrating it in the FRIA process as required in Section 4.2. For the tool to reuse a DPIA in this manner, the tool can either ask the user to input the DPIA outputs (information about the processing of personal data, risks identified, mitigations identified or applied, impact to rights) and then integrate it within the FRIA inputs, or use some developed interoperable specification or technical mechanism through which it can automatically extract the relevant information.

\subsection{Stage 3: Information gathering}
In the third stage, the tool helps with collecting the inputs defined in Section 3.2 required for the FRIA process, and helps with obligations in Art.27-1. The tool can vary from being completely manual (e.g. a form) or completely automated (e.g. extracting or inferring technical information from existing documents or pre-defined use-cases).

The input information required here relates to the AI system and its operations such as its intended purpose, relevant entities and their roles, involved data, and specific requirements for operation. The tool can either directly ask the user to provide this information, or use automation to derive the information. Based on the categories identified in Section 3.2, the following additional inputs are also needed.
\setlist{nolistsep}
\begin{enumerate}
    \item \textbf{Purpose compatibility:} an assessment of whether the purposes of the AI system being deployed are compatible with the intended purposes for which the AI system was developed and put on the market (Art.27-1a). The answer to this must include a binary value in the form of compatible or not-compatible. For internal compliance management purposes, the tool should also aim to collect additional information such as when this assessment was performed, by whom, whether the AI system has had any change since then, and recording links or references to any documentation that must be associated with this process. If the tool is designed to support performing an assessment of purpose compatibility, it can use approaches such as those defined in literature \cite{golpayeganiAICardsApplied2024} which provide a heuristic to flag situations requiring an assessment and provide a methodology to perform it based on specific information categories already available to the organisation.
    \item \textbf{Affected Persons:} the direct or indirect involvement of natural persons and groups of persons which is categorised based on whether they are the user/operator or subject of the AI system, whether they are intended or unintended, active or passive, informed or uninformed, and whether they have a relationship with the deployer - such as through a contract for service provision. In addition to this, information on what controls, if any, are available to such persons in the context of the AI system is also necessary - such as whether they can view the output, correct it, and can opt-in/out. It is also essential to consider whether such affected persons can be considered to be a `vulnerable group' by virtue of their nature (e.g. minors) or social vulnerability (e.g. belonging to a protected group). It is also essential to not only consider persons being identified as directly involved subjects, but also as potentially being excluded - which is important for determining impacts such as discrimination in later stages.
    \item\textbf{Personal Data and Special Category Personal Data:} a binary (yes/no) assessment of whether the AI system uses or produces or is capable of producing personal data, and whether such data belongs to special categories under GDPR, and whether any inputs or outputs of the AI system can be classified as profiling or decisions. It is also essential to enquire regarding the quality of data in terms of specific characteristics relevant to its role e.g. completeness for training, or accuracy for output and verification. This information can be collected directly by the tool or retrieved from a `datasheet' if present.
    \item\textbf{Technical and operational metrics:} pertinent details regarding the performance, robustness, and cybersecurity of the AI system (Anx.IV) that are necessary to identify risks in use of the AI system. This includes the ability to understand the operations and outputs (explainability) in a technical sense i.e. how was a particular result calculated, and organisational sense i.e. what processes and information were involved in producing the result.
    \item\textbf{Degree of control over the AI system:} an assessment regarding the ability of the deployer to control the AI system, such as to shut it down, change specific functions in a way that changes the output, provide different inputs, or make other modifications in a manner that materially changes the information required in a FRIA. This includes the ability to deploy specific measures in response to risks. Where the deployer does not have this ability, information on whether such measures have already been incorporated by the provider is necessary for the risk mitigation process in later stages. 
    \item\textbf{Dependencies:} identification of what resources (hardware, software, components) the AI system is dependant on to function, who controls it (e.g. a software that is operated by a third party which the deployer can only consume/use).
\end{enumerate}

Based on this as the input information, the tool then identifies information (either manually from the user or derives it automatically) on the specific risks associated with the use of AI system, the consequences of those risks, and whether this can lead to an impact on rights. Based on the existing body of work regarding the DPIA process which also includes a similar assessment of rights, we identify the following information for the tool to collect or infer regarding risks, mitigation measures, and consequences. 
\setlist{nolistsep}
\begin{enumerate}
    \item\textbf{Risks:} the system stops working completely, there is a component failure (where data is also considered as a component, and failure includes both operational failure e.g. component fails to function in operation and existential failure e.g. data is incomplete), system or component experiences reduced efficiency, system or component gives an incorrect output, there is an unauthorised use of the system, the system is attacked/damaged, the system is hacked/taken control of.
    \item\textbf{Risk Sources:} the system itself, a specific component of the system (with data being considered as a component), the user/operator of the system, the (human) subject of the system, the environment of use (e.g. public area), malicious actors.
    \item\textbf{Consequences:} reduction in service quality or availability, exclusion from service or process, loss of opportunity, delays in service or producing outputs, denial of a service - whether the use of AI system itself or of another service or process, unauthorised use, physical effect on the subject (including physical harm), psychological effect (including psychological distress, harms, etc.), cybersecurity incident (including data breach).
    \item\textbf{Mitigation Measures:} prevent or reduce the likelihood or the severity of the risk or consequence event from occurring, establish monitoring controls to identify risk and the correct and continued operation of mitigation measures - including human oversight measures, performing technical and organisational audits to establish robustness and trustworthiness of processes, conducting organisational training, providing literacy and awareness to specific stakeholders.
\end{enumerate}

Using these categories, the tool can assist in identifying the risks that are relevant to the AI system, when they can arise, what their consequences will be for the affected persons, whether they are significant in nature based on the use-case, whether their effects are temporary or lasting, and how they can be addressed through different mitigation measures. For example, for an automated passport checking system, the tool can help identify risks such as when can the system stop detecting passports (i.e. the system shuts down), what could be the causes of this (e.g. electrical problems), what this might mean for specific persons (e.g. delays for passengers, no immigration service provided), and what can be done about it (e.g. a human officer manually performs passport verification). Based on this, the tool can assist in determining the residual risk level e.g. through a risk matrix, and assessing whether the risk level is acceptable for the use-case or not. 

\subsection{Stage 4: Producing outputs} In the fourth stage, the tool helps with producing the outputs from the FRIA process - namely the risks of harmful impacts on health, safety, and fundamental rights, and helps with obligations in Art.27-1. The inputs for the tool consist of the information described in Section 3.2, and the operation of the tool can vary from being completely manual (e.g. asking the user to identify the impacts) to being completely automated (e.g. inferring impacts from given information).

In order to assist with identifying impacts on specific fundamental rights, the tool can utilise existing frameworks such as the ALIGNER FRIA template \cite{FundamentalRightsImpact} which identifies a risk level from inputs to represent the risk to fundamental rights and freedoms. Or it can let the user perform the assessment by providing a structured method that takes each of the consequences identified in Stage 3 and explores what is the implication if it were to occur by taking in to account the affected person and the intended purpose of the AI system, and explores whether this has a bearing on a particular fundamental right, and if so - then what kind of an impact it is. 

For example, in the earlier example of an automated passport control system, if a person is excluded from the use of this service due to the AI system not functioning accurately for that person's racial or other biological characteristics, the consequences to be identified can include exclusion from service, temporal detriment, or psychological distress, which then need to be identified as being of significance i.e. as a `legal effect'. Were this to occur, then the right to non-discrimination in Art.21 of the Charter of Fundamental Rights is of relevance here.

To understand what kind of impact this is, it is necessary to consider whether the right is active i.e. it requires an exercise, or passive i.e. it always applies, and whether the right is absolute i.e. they must always be provided and respected, or limited i.e. it can be restricted or overridden in certain situations. In addition, we also propose the following (mutually non-exclusive) impact categories for rights:
\setlist{nolistsep}
\begin{enumerate}
    \item The right is violated i.e. there is an infringement or breach that constitutes a `violation';
    \item The exercise of the right is prevented;
    \item The right is limited i.e. there is a limitation or restrictions on the scope or exercise of that right;
    \item The right is denied i.e. the existence or applicability of that right is denied or to be applicable;
    \item The right is unfulfilled i.e. there is a failure to meet or complete the requirements;
    \item The right is obstructed i.e. there is an interference in the exercise of the right with the intention or consequence of preventing its exercise;
\end{enumerate}
By using these categories, the remedial measure for each particular impact to the right can be identified. For example, if the passenger in the passport control situation is denied the use of a different AI system or a human replacement, this can count as a limitation of the freedom of free movement and of the right to non-discrimination, and if left unresolved also as a violation of those rights and freedoms. The remedial measure in this case corresponds to the category of impact i.e. the limitation of a right requires either finding a way to remove the limitation or to provide an alternative mechanism where the limitation is not present. This means establishing  procedures that identify when the AI system can fail or produce inaccurate outputs based on racial and biological features (i.e. through bias), finding ways to reduce and mitigate these, and establishing manual oversight and intervention procedures to ensure people have an alternative mechanism to fulfil the legally required immigration process where the automated passport control system cannot function appropriately.

\subsection{Stage 5: Notifying authorities} In the fifth stage, the tool helps with the notification of the FRIA to the authorities as required in Art.27-3. The input to this tool will be a completed FRIA along with relevant metadata such as timestamps and entity information, and the output could consist of an acknowledgement from the authority including timestamps and identifiers. The tool could assist here in creating the documentation to be sent in the notification, or can also act as a communication tool to directly notify the authority with the relevant information. As before, the tool can do these operations manually (i.e. the user decides when and what to notify) or automatically (i.e. the tool automatically sends notifications for specific conditions).

\section{Conclusion \& Future Work}
In this article, we interpreted the GDPR's DPIA and the AI Act's FRIA as processes involving information for inputs and outputs. Based on this, we identified the commonality in both to identify where a DPIA might be reused to create a FRIA, or based how a DPIA and a FRIA could be developed concurrently. We then used this understanding to express the FRIA as a 5-step process, including the reuse of a DPIA and the notification obligation. Through this, we have outlined a methodological approach for the implementation of the FRIA questionnaire and automated tool, as required in AI Act Article 27, and provided the first discussions on how and where such an automated tool could be used, and what information it would require, and the varying uses of automation in the FRIA process. Our work has significance in the developing interpretations of the AI Act, particularly regarding the novel clauses associated with FRIA, and provides a practical approach for the information and documentation that is needed along with how technological tools can support in tasks associated with it.

The work presented in this article represents the preliminary stages of a larger approach towards reusing DPIAs in FRIA, and creating automated tools to support the FRIA processes. Both the DPIA and the FRIA are complex legal processes which require varying information depending on the use-case involved, and therefore will also require information management tools that can support it. We therefore identify the necessity to develop approaches that facilitate the representation of information associated with DPIA and FRIA processes in a structured, machine-readable, and interoperable manner so as to enable the creation of useful tools and information services that can work across use-cases and for different stakeholders. We plan to use and extend the Data Privacy Vocabulary (DPV) \cite{panditDataPrivacyVocabulary2024} for representing DPIA and FRIA information as the DPV enables representing information for both GDPR and AI Act, and provides a large collection of taxonomies that aid representing real-world use-cases. 

We also identify the need to further investigate the requirements for conducting DPIA and FRIA based on the legal obligations defined in GDPR and AI Act respectively, as these are crucial requirements that form the first stage in both assessments. For this, we aim to represent the information required to assess whether a DPIA and a FRIA is needed using the DPV, and to then use this as the basis to develop a tool that can automate the assessment for when a DPIA or a FRIA is needed, and which can be further enhanced to support the individual in expressing their use-case and discovering risks and impacts through logical reasoning methods.

\begin{acknowledgments}
This work was funded by the ADAPT SFI Centre for Digital Media Technology, which is funded by Science Foundation Ireland through the SFI Research Centres Programme and is co-funded under the European Regional Development Fund (ERDF) through Grant\#13/RC/2106\_P2.
\end{acknowledgments}

\bibliography{references}

\end{document}